%% file: cosine100modulation.tex
\documentclass[aps,prl,reprint,superscriptaddress,showkeys,showpacs]{revtex4-1}

\usepackage{graphicx} %
\usepackage{amsmath} %
\usepackage{amssymb} %
\usepackage{subfigure} %
\usepackage{float} %
\usepackage{upgreek} %
\usepackage{multirow} %
\usepackage{color} %
\usepackage{lineno} %
\usepackage{hyperref} %
\usepackage{booktabs}
\usepackage{placeins}
\usepackage[normalem]{ulem}
\usepackage[usenames,dvipsnames]{xcolor}
\hypersetup{colorlinks=true, urlcolor=blue, linkcolor=blue, citecolor=red} %
\usepackage{lineno}
\usepackage{lipsum}
\usepackage{tabularx}

\newcommand{\thisbegindate}{October 21, 2016}
\newcommand{\thisenddate}{July 18, 2018}
\newcommand{\totaluptime}{1.7\,years}
\newcommand{\totalexposure}{97.7\,kg$\cdot$years}
\newcommand{\totalmass}{61.3\,kg}
\newcommand{\bestfitamplitdude}{0.0092$\pm$0.0067}
\newcommand{\bestfitphase}{127.2$\pm$45.9}

\makeatletter
\renewcommand*{\@fnsymbol}[1]{\ensuremath{\ifcase#1\or \dagger\or *\or \ddagger\or
    \mathsection\or \mathparagraph\or \|\or **\or \dagger\dagger
    \or \ddagger\ddagger \else\@ctrerr\fi}}
\makeatother

\begin{document}

\title{Search for a Dark Matter-Induced Annual Modulation Signal in NaI(Tl) with the COSINE-100 Experiment}

\input{COSINE-100_authors_March2019.tex}
\date{\today}

\begin{abstract}

We present new constraints on the dark matter-induced annual modulation signal using \totaluptime\, of COSINE-100 data with a total exposure of \totalexposure. The COSINE-100 experiment, consisting of 106\,kg of NaI(Tl) target material, is designed to carry out a model-independent test of DAMA/LIBRA's claim of WIMP discovery by searching for the same annual modulation signal using the same NaI(Tl) target. The crystal data show a 2.7 cpd/kg/keV background rate on average in the 2--6 keV energy region of interest. Using a $\chi$-squared minimization method we observe best fit values for modulation amplitude and phase of \bestfitamplitdude\, cpd/kg/keV and \bestfitphase\,d, respectively.


\end{abstract}

\maketitle



Cosmological observations give strong evidence that 27\% of the energy content of the Universe exists in the form of nonluminous dark matter~\cite{Ade:2013zuv}, unaccounted for by the standard model of particle physics~\cite{pdg2018}. One theoretically favored model of dark matter posits the existence of weakly interacting massive particles (WIMPs)~\cite{Lee,Goodman} that interact only through the gravitational and weak scale forces and have a mass on the GeV to TeV scale~\cite{WIMP,Bertone}.
Within the context of the standard halo model, there will be an annual modulation in the dark matter--nucleon interaction rate with a period of one year~\cite{Colloquium,Freese:1987wu,Lewin:1996}. One experiment, DAMA/LIBRA, observes annual modulations in the detected event rate with a significance exceeding 12\,$\sigma$, which they attribute to the presence of dark matter~\cite{Bernabei:2005hj,Bernabei:2013xsa,DAMAPhase2}. DAMA/LIBRA’s observation is inconsistent with other experiments under most well-motivated WIMP dark matter models~\cite{LUX2017,XENON2018,PandaXII2017,CDMS,CRESST,PICO,KIMS,SIMPLE,XMASS2018}; however, none of these other experiments have used the same target material as DAMA, thallium-doped sodium iodide [NaI(Tl)] scintillating crystals. Thus, these comparisons are necessarily dependent on the particular model of WIMP-nucleus scattering and the assumed WIMP halo structure. 

The COSINE-100 experiment aims to resolve this tension in the field by performing a model-independent test of DAMA's observation using the same detector material, NaI(Tl), as DAMA. Previously, we have performed a model-dependent test of DAMA and found that DAMA's observed annual modulation cannot be explained by spin-independent WIMP-nucleus scattering in the context of the standard halo model~\cite{COSINE_WIMP}. Additionally, there are several other experiments aimed at performing model-independent tests of DAMA, including DM-Ice17~\cite{BarbosaDeSouza:2017}, KIMS~\cite{KIMS_NaI}, SABRE~\cite{SABRE}, and ANAIS-112~\cite{ANAIS2019EPJC,ANAIS2019}, which has recently reported its first result.

COSINE-100 is located at the Yangyang Underground Laboratory (Y2L) in South Korea, with $>$700\,m of rock overburden. It consists of eight NaI(Tl) crystals with a total mass of 106\,kg immersed in 2200\,l of liquid scintillator (LS) that reduces internal and external backgrounds~\cite{KIMS_LS}. Each NaI(Tl) crystal is optically coupled to two photomultiplier tubes (PMTs), each of which detects scintillation photons with the signals recorded as 8\,$\mu$s waveforms~\cite{COSINE_daq}. These eight crystals are referred to as Crystal 1 (C1) to Crystal 8 (C8). C1, C5, and C8 are excluded from this analysis due to their high background (about twice that of the other crystals), high noise rate (C1), and low light yield (C5 and C8), for a total effective mass of \totalmass. The detector is surrounded by passive and active shielding that includes, from the inside out, copper plates of 3\,cm in total thickness, 20\,cm of lead, and 3\,cm of 37 plastic scintillator panels for cosmic ray muon tagging~\cite{COSINE_muon}. More details of the experimental apparatus are presented in Ref.~\cite{COSINE_detector}.

Data taking for COSINE-100 began in September 2016, and the analysis presented here covers an exposure of \totaluptime\ years, spanning from \thisbegindate\ to \thisenddate. Several datasets from C2 and C7 are excluded due to excessive noise levels. The total exposure used in this analysis corresponds to \totalexposure.  

\begin{figure}
	\centering
	\includegraphics[width=\columnwidth]{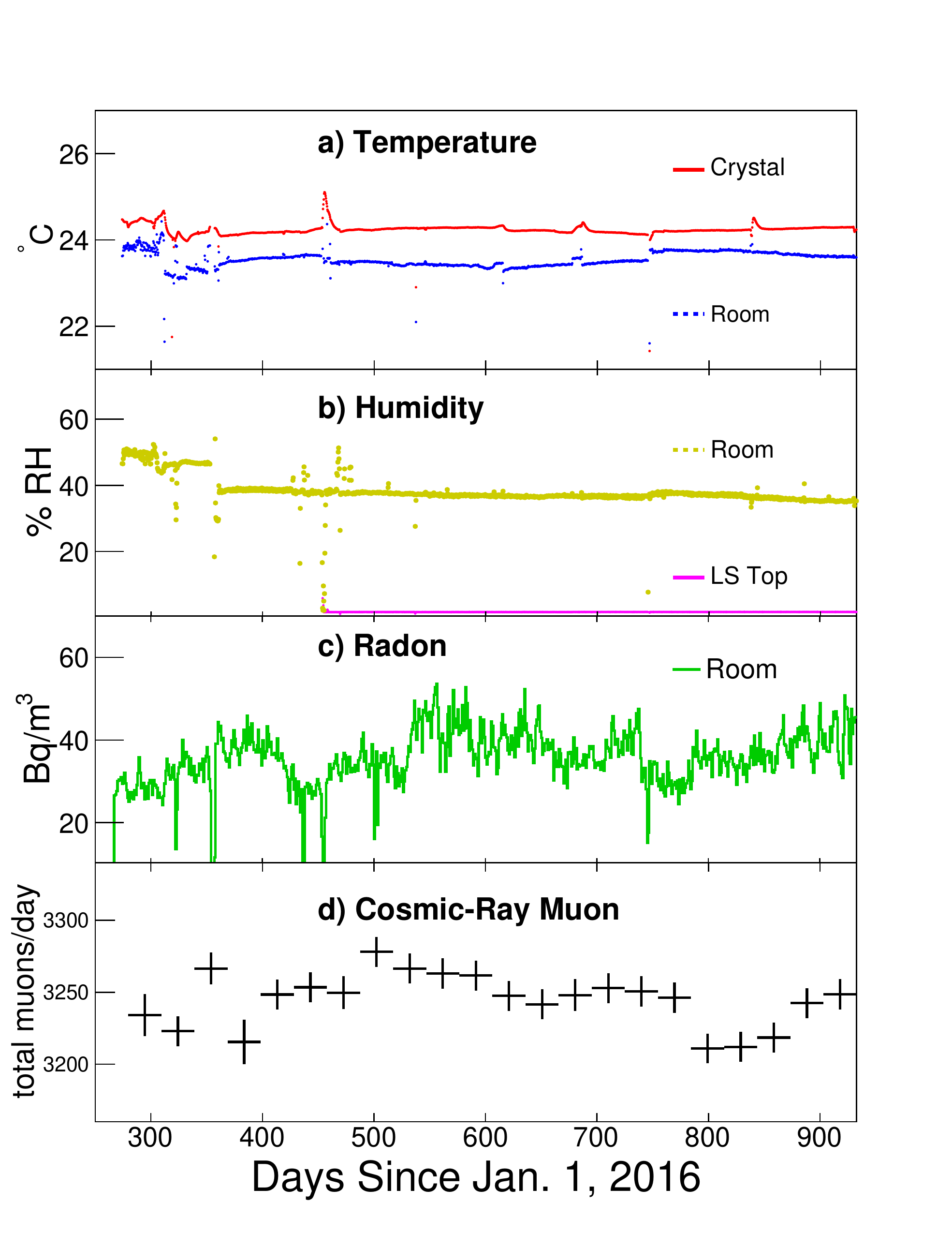}
	\caption{COSINE-100's environmental parameters as a function of time. (a) Detector room and near-crystal temperature. (b) Relative humidity for the detector room and the top volume of acrylic box, at the top of the LS. 
	Note that the measurement taken at the top of the LS began on day 450. (c) The radon level in the detector room air. (d) Rate of muons passing through the detector over time. Here, the rate is binned in 30-day intervals. }
	\label{fig:env_monitor}
\end{figure}

The overall stability of the detector is closely monitored to ensure that neither environmental nor detector effects can create an artificial dark matter signal~\cite{COSINE_detector}. Humidity and temperature of the detector room are maintained at 40.0$\pm$3\%\,RH (relative humidity) and 23.5$\pm$0.3\,$^{\circ}$C, respectively. Gas boiloff from liquid nitrogen is introduced into the space above the liquid scintillator inside the inner copper chamber at a rate of 3 l/min to purge radon and prevent contact between the LS and oxygen or water vapor, which maintains a high scintillator light yield. The humidity inside the shielding structure is kept at $<$\,5\%\,RH and the high heat capacity helps to keep the temperature within the liquid stable at 24.2$\pm$0.1\,$^{\circ}$C. The radon level in the detector room is measured at 36$\pm$10 Bq/m$^{3}$. The time dependence of temperature, humidity, radon, and cosmic ray muons~\cite{COSINE_muon} is shown in Fig.~\ref{fig:env_monitor}. The spikes in Fig.~\ref{fig:env_monitor}(a) are due to power outages or air conditioning failures; these periods are excluded from the data.  The effects of temperature and radon level on the pulse shape, light yield, and overall performance of the NaI(Tl) detectors and of the full detector were reported in Ref.~\cite{Schneid1977}. A monitoring of fast neutrons inside the detector room has recently begun in Summer 2018~\cite{COSINE_neutron}.

The gain of the PMTs is monitored by measuring the position of the 46.5\,keV peak from $^{210}$Pb decays that occur in the NaI(Tl) crystal bulk. The gain is tracked and modeled as a piecewise linear function in time. Observed gain shifts over time are corrected for in each PMT. After correction, the 46.5\,keV peak is stable to within 0.1\% on average. We assess the efficacy of this gain correction method within the 2--6\,keV region of interest by measuring the position of the 3.2\,keV decay peak from $^{40}$K over time; the position of the decay peak is stable to within $<$\,2\% on average in the dataset used in the analysis. 

\begin{figure}
	\centering
	\includegraphics[width=0.9\columnwidth]{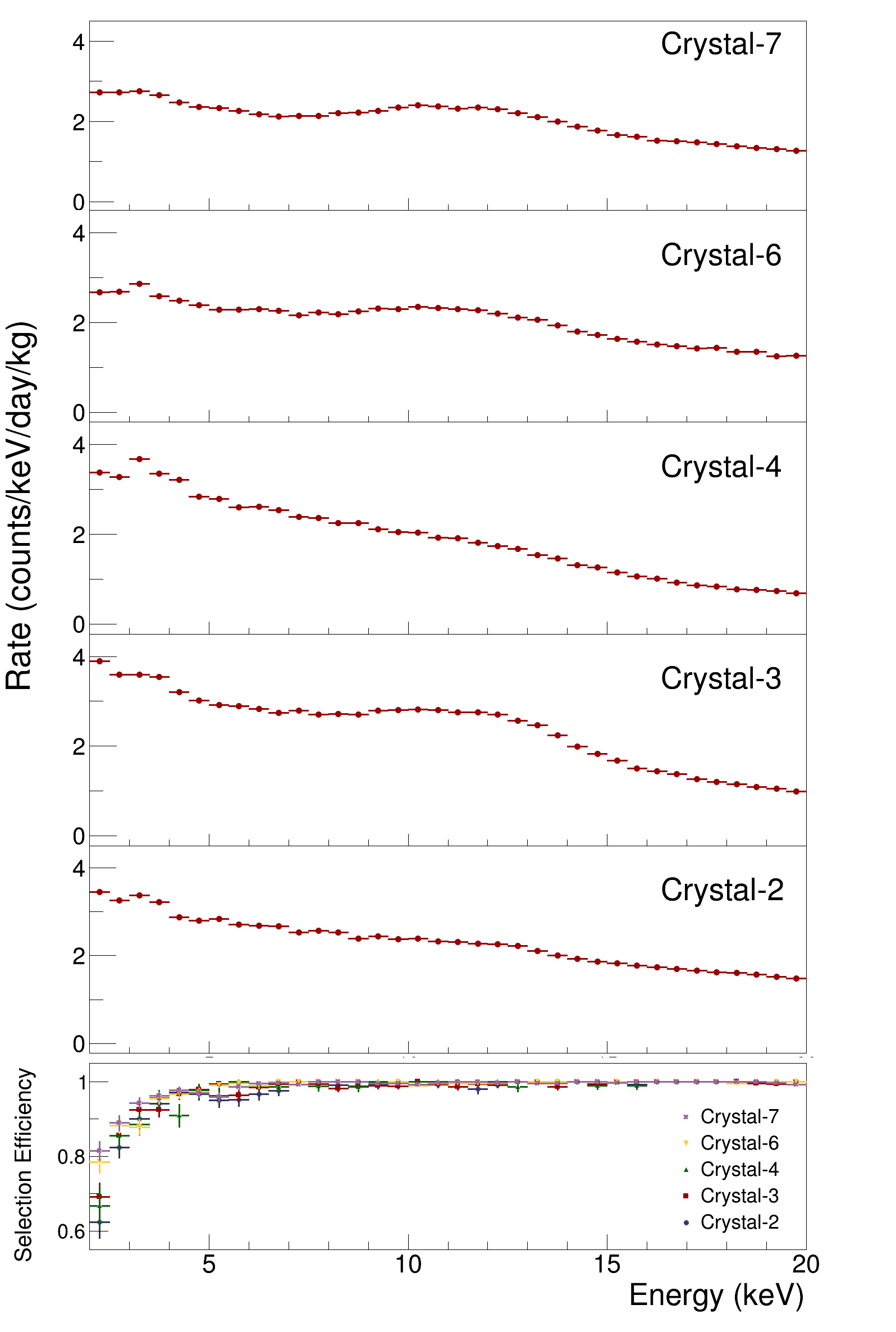}
    \caption{Efficiency-corrected and time-integrated energy spectra for the five crystals used in this analysis between 2--20\,keV (top panels) and signal selection efficiency evaluated using $^{60}$Co calibration data (bottom panel). The efficiencies at 2\,keV are $>$60\% for all crystals. The primary sources of background in the crystals are $^{210}$Pb and $^{40}$K, which are lower for Crystal 6 and Crystal 7. These spectra are obtained using the full dataset considered in this analysis.}
    \label{fig:eff_spectrum}
\end{figure}

Events that trigger more than one crystal, pulses with pulse shapes that are inconsistent with a NaI(Tl) scintillation signal, e.g.,\ PMT related noise, are rejected~\cite{KIMS_NaI,COSINE_detector,ANAIS2014}. We use two boosted decision trees, which are multivariate analysis algorithms (BDTs)~\cite{BDT}, to remove PMT-related and other noise events, which we call BDT1 and BDT2. BDT1 is used to remove PMT-induced noise and is based on the amplitude-weighted average time of a pulse, the ratios of the leading- and trailing-edge charge sums relative to total charge, and the difference of deposited charges between the two PMTs~\cite{COSINE_EFT}. It is trained with a sample of signal-rich, energy-weighted events from a $^{60}$Co calibration run for signal, and single-hit events from the WIMP-search physics-run data for noise, with the latter mostly triggered by PMT noise events. The second BDT, BDT2, includes weighted higher-order time moments and eliminates intermittent PMT discharge-triggered events that have slower pulse decay times. The event selection technique and criteria are described more in detail in Refs.~\cite{COSINE_detector,COSINE_WIMP}.

The same BDT selections were applied to the Compton-scattered low energy events from a $^{60}$Co calibration run to estimate the event selection efficiency. The efficiency is the ratio of events that survive the selection to the total number of signal events. Uncertainties on the efficiency follow binomial statistics. Figure~\ref{fig:eff_spectrum} shows the event selection efficiency as a function of energy, along with the efficiency-corrected, 2--20\,keV spectra of the five crystals used in this analysis. The spectra are well modeled with a \textsc{GEANT}4-based simulation~\cite{geant, geant_2, COSINE_background}; the 3.2\,keV $^{40}$K peak is clearly visible in C2 and C4, whereas the overall background levels in C6 and C7 are lower than in other crystals because of their lower $^{210}$Pb and $^{40}$K contamination levels.


\begin{figure}
	\centering
	\includegraphics[width=\columnwidth]{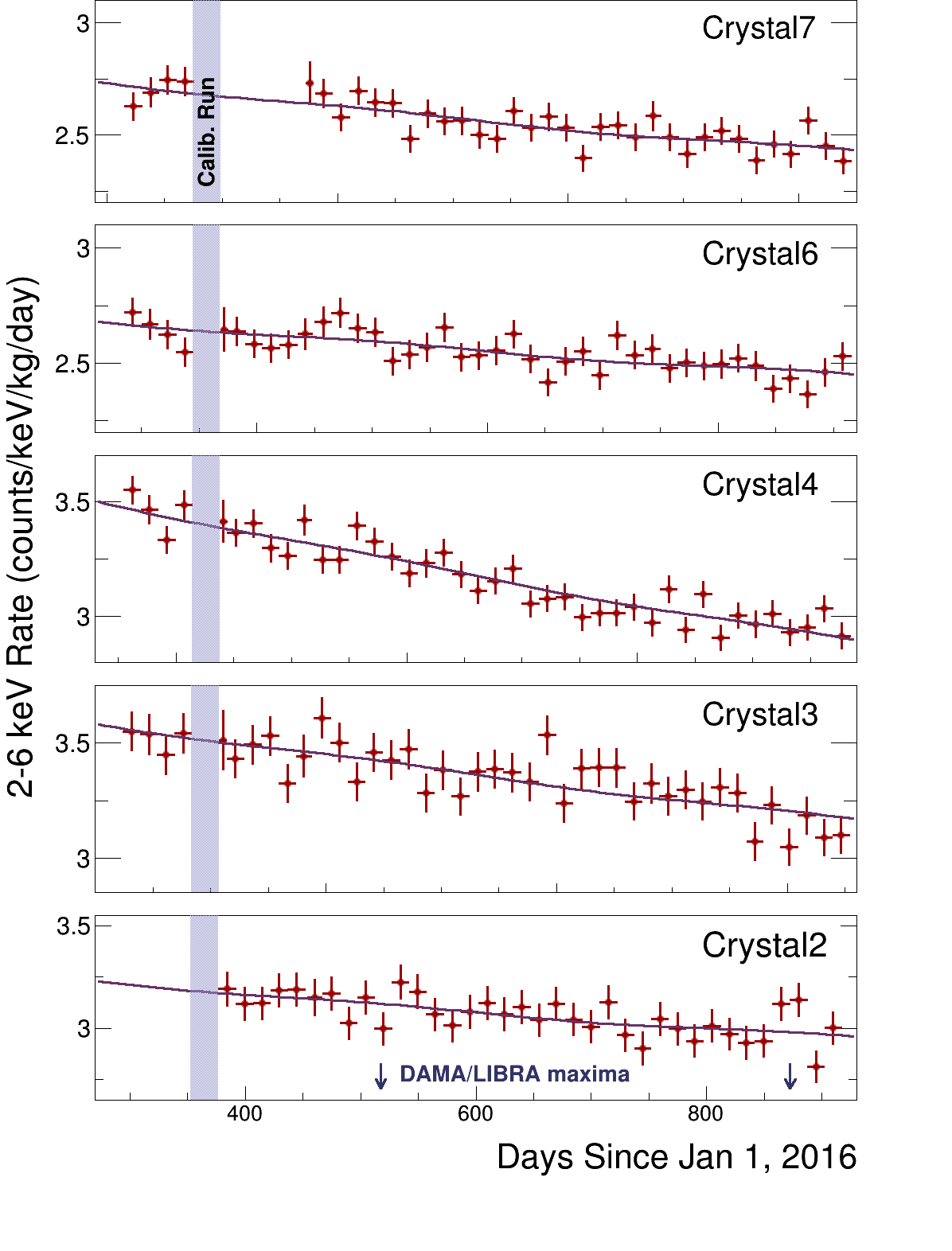}
	\caption{Rate vs.\ time for Crystals 2, 3, 4, 6, and 7 from \thisbegindate\ to \thisenddate\ for the 2--6\,keV energy region binned in 15-day intervals. The histograms show the result of the fit described in the text. Solid blue arrows indicate the peak date in the modulation as reported by DAMA/LIBRA~\cite{DAMAPhase2}. Data taking was suspended for calibrations at the end of 2016 as indicated by the shaded region.}
    \label{fig:rate_versus_time}
\end{figure}

In order to confirm our background understanding and account for possible systematic effects that could appear over time, we investigated a control sample of multiple-hit events in the 2--20\,keV energy region with statistics comparable to that in the region of interest (ROI) of 2--6\,keV. These are events in which multiple NaI(Tl) crystals are triggered or a single crystal is triggered along with the LS and, thus, cannot be attributable to typical WIMP dark matter interactions. They comprise 20\% of the total signal event sample.

We also consider the possibility that certain event types that are removed during event selection could cause a modulation signal. The noise events observed in the COSINE-100 detector are systematically categorized and studied to understand how their removal affects the signal region counting rate over time. This study confirmed none of the cut individually show a modulation in the removed events and have negligible impact on the modulation of signal events.


The event rates as functions of time are modeled as:
\begin{equation}\label{eq:fit}
	\textrm{Rate} = C+ p_{0}\exp{(-\frac{ln2t}{p_{1}})} + A\cos{\frac{2\pi(t-t_{0})}{T}},
\end{equation}
where $C$ is a constant offset constrained by background modeling as described in Ref.~\cite{COSINE_background}, and $p_{0}$ and $p_{1}$ are the amplitude and half-life for an exponentially decaying background, which models cosmogenically activated backgrounds. The modulation is described by $A$, $T=365.25$ d, and $t_{0}$, its amplitude, period, and phase, respectively.

\begin{figure}[]
	\centering
	\includegraphics[width=\columnwidth]{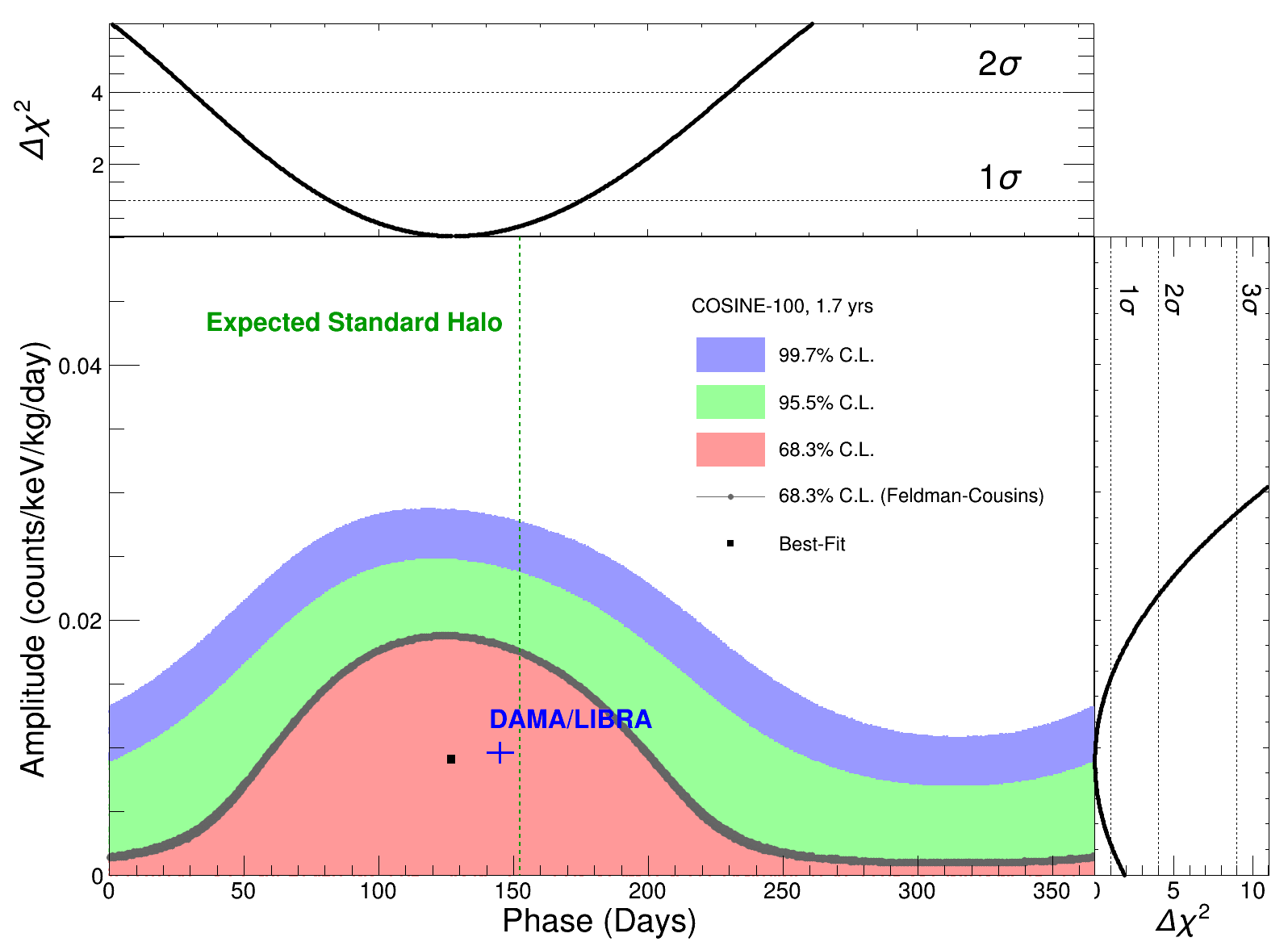}
	\caption{The COSINE-100 best fit and 68.3\%, 95.5\%, and 99.7\% C.L. contours as functions of modulation amplitude and phase relative to January 1, for a fixed period of 365.25 d. A Feldman-Cousins technique is used as a crosscheck and resultant 68.3\% C.L. is shown. The amplitude and phase reported by DAMA/LIBRA in the 2--6\,keV energy interval with statistical uncertainties (blue cross) and the phase expected from the standard halo model (June 2) are overlaid for comparison. Top and side panels show the dependence of $\Delta\chi^{2}$ on phase and amplitude, respectively, along with two-sided significance levels.}
    \label{fig:chisq2D}
\end{figure}

The data from all crystals were fit simultaneously with the same amplitude and phase amongst all crystals but allowing for different exponential decaying and constant background components to account for the varying background levels across different crystals. Figure~\ref{fig:rate_versus_time} shows the COSINE-100 event rates over time for the 2--6\,keV ROT in the crystals used in this analysis, where recorded 670 events/day on average, i.e.\ 2.7 cpd/kg/keV. We performed $\chi$-squared minimization fits for the modulation amplitude with the period fixed at 365.25 d with the phase as a free parameter and, also, with it fixed at the halo-model expectation value of 152.5 d and the DAMA/LIBRA-observed value of 145 d. Initially, we performed a blinded analysis by only analyzing $\sim$9\% of the data, evenly distributed in time. However, during unblinding, we observed a large number of anomalous noise events within the signal region. This led us to develop BDT2 in order to remove these anomolous events and to reanalyze the data unblinded. The best fit to the 2--6\,keV range has a modulation amplitude of \bestfitamplitdude\,cpd/kg/keV with a phase of \bestfitphase\,d. A log-likelihood parameter estimation of the annual modulation with amplitude and phase as free parameters shows that the current data from COSINE-100 is consistent with both the DAMA/LIBRA annual modulation result and the null hypothesis of no modulation at the 68.3\% C.L. as shown in Fig~\ref{fig:chisq2D}. A Feldman-Cousins method~\cite{Feldman} was also used to crosscheck the result, and returned a consistent C.L.

Table~\ref{tab:resultNumbers} summarizes the result of the various fitting scenarios used for the 2--6\,keV energy interval. The period is fixed at 365.25 d (one year) for all scenarios, whereas the phase is either floated freely or fixed at 152.5 d as expected from the standard halo model. COSINE-100 is the only NaI(Tl) experiment with a LS veto surrounding the crystals providing additional capabilities for rejection of external background. As a crosscheck, we show the annual modulation fit results both with and without the LS veto. The LS veto removes backgrounds and improves the uncertainties on the annual modulation amplitudes by 4\%.

\begin{table*}[]
	\centering
	\caption{Summary of fit results for the modulation and null hypotheses for the 2--6\,keV energy region in COSINE-100. Detector rates were fit to Eq.~\eqref{eq:fit}, with the period fixed at 365.25 d. Results with phase floated and fixed at 152.5 d are listed. The result without using the LS veto is presented as a crosscheck. DAMA/LIBRA results~\cite{DAMAPhase2} and the ANAIS-112 2019 result~\cite{ANAIS2019} are also shown. 
	}
	\begin{tabular}{lccccc}
		\hline
		\hline
		Configuration& $\chi^{2}$ & DOF & \textit{p}-value & Amplitude (cpd/kg/keV) & Phase (d)\\
		\hline
		COSINE-100   & 175.3 & 174 & 0.457 & 0.0092$\pm$0.0067 & 127.2$\pm$45.9\\
		DAMA/LIBRA (Phase1+Phase2) &  $\cdot\cdot\cdot$ & $\cdot\cdot\cdot$  & $\cdot\cdot\cdot$ & 0.0096$\pm$0.0008 & 145$\pm$5\\
		\hline
		COSINE-100  & 175.6 & 175 & 0.473 & 0.0083$\pm$0.0068 & 152.5 (fixed)\\
		COSINE-100 (Without LS)  & 194.7 & 175 & 0.147 & 0.0024$\pm$0.0071 & 152.5 (fixed)\\
		ANAIS-112                       & 48.0  & 53   & 0.67  & -0.0044$\pm$0.0058 & 152.5 (fixed)\\
		DAMA/LIBRA (Phase1+Phase2)      & 71.8  & 101  & 0.988 & 0.0095$\pm$0.0008 & 152.5 (fixed)\\
		\hline
		\hline
	\end{tabular}
	\label{tab:resultNumbers}
\end{table*}


\begin{figure}
	\centering
	\includegraphics[width=\columnwidth]{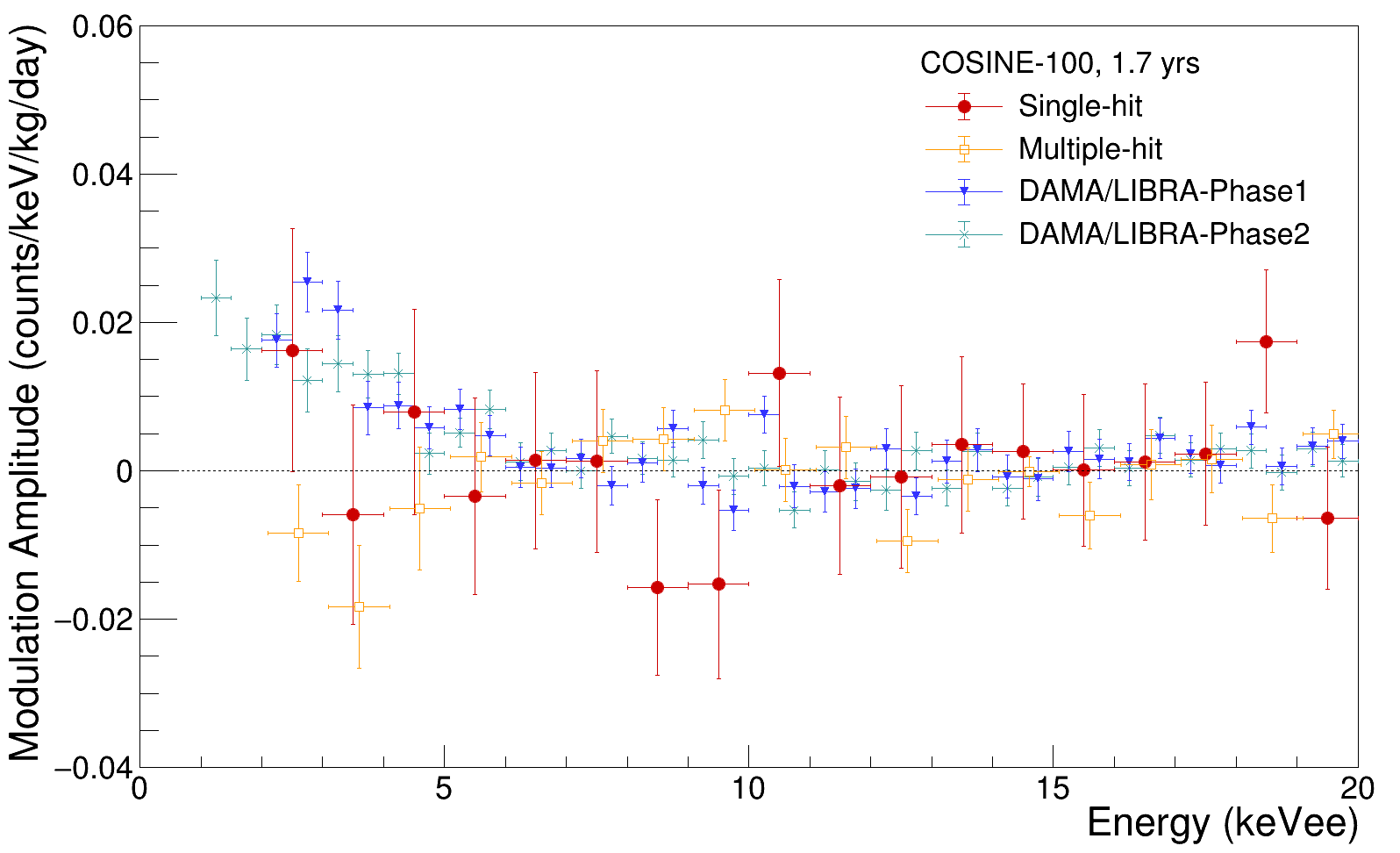}
	\caption{Modulation amplitude as a function of energy in 1\,keV bins for the 1.7 year COSINE-100 single-hit (red closed circle) and multiple-hit (orange open circle) events. DAMA/LIBRA phase 1 (blue) and phase 2 (green) from Ref.~\cite{DAMAPhase2} are also shown for reference. The period and phase are fixed at 365.25 d and 152.5 d. Horizontal error bars represent the width of the energy bins used for the analysis. Vertical error bars are $\pm 1\sigma$ errors on the binned modulation fit amplitudes.}
    \label{fig:modulationVsEnergy}
\end{figure}

The best fit modulation amplitudes as a function of energy with 1\,keV energy bins are shown in Fig.~\ref{fig:modulationVsEnergy}. These fits were performed with a fixed period of one year and the phase fixed at 152.5 d. 

In summary, we report the results from the search for a dark matter--induced annual modulation signal in NaI(Tl) based on 1.7 years of COSINE-100 data. A fit to the 2--6 keV energy range returns a modulation amplitude of \bestfitamplitdude\,cpd/kg/keV with a phase of \bestfitphase\,d. At 68.3\% C.L., this result is consistent with both the null hypothesis and DAMA/LIBRA's 2--6\,keV best fit value. We expect COSINE-100 will attain 3\,$\sigma$ coverage of DAMA region within five years of data exposure. Future searches with \mbox{COSINE-100} will utilize a larger dataset and lower energy threshold of at least 1\,keV with improved event selection efficiency and are expected to reduce the required exposure for 3\,$\sigma$ coverage.  


\begin{acknowledgments}
We thank the Korea Hydro and Nuclear Power (KHNP) Company for providing underground laboratory space at Yangyang. This work is supported by: the Institute for Basic Science (IBS) under project code IBS- R016-A1 and NRF-2016R1A2B3008343, Republic of Korea; UIUC campus research board, the Alfred P. Sloan Foundation Fellowship, NSF Grants No. PHY-1151795, No. PHY-1457995, and No. DGE-1122492, WIPAC, the Wisconsin Alumni Research Foundation, United States; STFC Grants No. ST/N000277/1 and No. ST/K001337/1, United Kingdom; and Grant No. 2017/02952-0 FAPESP, CAPES Finance Code 001, Brazil. We thank P.T.~Surukuchi for helpful discussions.
\end{acknowledgments}

\bibliography{cosine100modulation}

\end{document}

%% file: COSINE-100_authors_March2019.tex
\author{G.~Adhikari}
\affiliation{Department of Physics, Sejong University, Seoul 05006, Republic of Korea}
\author{P.~Adhikari}
\thanks{Present address: Department of Physics, Carleton University, Ottawa, Ontario, K1S 5B6, Canada}
\affiliation{Department of Physics, Sejong University, Seoul 05006, Republic of Korea}
\author{E.~Barbosa~de~Souza}
\affiliation{Department of Physics and Wright Laboratory, Yale University, New Haven, Connecticut 06520, USA}
\author{N.~Carlin}
\affiliation{Physics Institute, University of S\~{a}o Paulo, 05508-090, S\~{a}o Paulo, Brazil}
\author{S.~Choi}
\affiliation{Department of Physics and Astronomy, Seoul National University, Seoul 08826, Republic of Korea} 
\author{M.~Djamal}
\affiliation{Department of Physics, Bandung Institute of Technology, Bandung 40132, Indonesia}
\author{A.~C.~Ezeribe}
\affiliation{Department of Physics and Astronomy, University of Sheffield, Sheffield S3 7RH, United Kingdom}
\author{C.~Ha}
\affiliation{Center for Underground Physics, Institute for Basic Science (IBS), Daejeon 34126, Republic of Korea}
\author{I.~S.~Hahn}
\affiliation{Department of Science Education, Ewha Womans University, Seoul 03760, Republic of Korea} 
\author{E.~J.~Jeon}
\affiliation{Center for Underground Physics, Institute for Basic Science (IBS), Daejeon 34126, Republic of Korea}
\author{J.~H.~Jo}
\thanks{Corresponding author}
\email{jayhyun.jo@yale.edu}
\affiliation{Department of Physics and Wright Laboratory, Yale University, New Haven, CT 06520, USA}
\author{H.~W.~Joo}
\affiliation{Department of Physics and Astronomy, Seoul National University, Seoul 08826, Republic of Korea}
\author{W.~G.~Kang}
\affiliation{Center for Underground Physics, Institute for Basic Science (IBS), Daejeon 34126, Republic of Korea}
\author{W.~Kang}
\affiliation{Department of Physics, Sungkyunkwan University, Suwon 16419, Republic of Korea}
\author{M.~Kauer}
\affiliation{Department of Physics and Wisconsin IceCube Particle Astrophysics Center, University of Wisconsin-Madison, Madison, Wisconsin 53706, USA}
\author{G.~S.~Kim}
\affiliation{Department of Physics, Kyungpook National University, Daegu 41566, Republic of Korea}
\author{H.~Kim}
\affiliation{Center for Underground Physics, Institute for Basic Science (IBS), Daejeon 34126, Republic of Korea}
\author{H.~J.~Kim}
\affiliation{Department of Physics, Kyungpook National University, Daegu 41566, Republic of Korea}
\author{K.~W.~Kim}
\affiliation{Center for Underground Physics, Institute for Basic Science (IBS), Daejeon 34126, Republic of Korea}
\author{N.~Y.~Kim}
\affiliation{Center for Underground Physics, Institute for Basic Science (IBS), Daejeon 34126, Republic of Korea}
\author{S.~K.~Kim}
\affiliation{Department of Physics and Astronomy, Seoul National University, Seoul 08826, Republic of Korea}
\author{Y.~D.~Kim}
\affiliation{Center for Underground Physics, Institute for Basic Science (IBS), Daejeon 34126, Republic of Korea}
\affiliation{Department of Physics, Sejong University, Seoul 05006, Republic of Korea}
\affiliation{IBS School, University of Science and Technology (UST), Daejeon 34113, Republic of Korea}
\author{Y.~H.~Kim}
\affiliation{Center for Underground Physics, Institute for Basic Science (IBS), Daejeon 34126, Republic of Korea}
\affiliation{Korea Research Institute of Standards and Science, Daejeon 34113, Republic of Korea}
\affiliation{IBS School, University of Science and Technology (UST), Daejeon 34113, Republic of Korea}
\author{Y.~J.~Ko}
\affiliation{Center for Underground Physics, Institute for Basic Science (IBS), Daejeon 34126, Republic of Korea}
\author{V.~A.~Kudryavtsev}
\affiliation{Department of Physics and Astronomy, University of Sheffield, Sheffield S3 7RH, United Kingdom}
\author{H.~S.~Lee}
\affiliation{Center for Underground Physics, Institute for Basic Science (IBS), Daejeon 34126, Republic of Korea}
\affiliation{IBS School, University of Science and Technology (UST), Daejeon 34113, Republic of Korea}
\author{J.~Lee}
\affiliation{Center for Underground Physics, Institute for Basic Science (IBS), Daejeon 34126, Republic of Korea}
\author{J.~Y.~Lee}
\affiliation{Department of Physics, Kyungpook National University, Daegu 41566, Republic of Korea}
\author{M.~H.~Lee}
\affiliation{Center for Underground Physics, Institute for Basic Science (IBS), Daejeon 34126, Republic of Korea}
\affiliation{IBS School, University of Science and Technology (UST), Daejeon 34113, Republic of Korea}
\author{D.~S.~Leonard}
\affiliation{Center for Underground Physics, Institute for Basic Science (IBS), Daejeon 34126, Republic of Korea}
\author{W.~A.~Lynch}
\affiliation{Department of Physics and Astronomy, University of Sheffield, Sheffield S3 7RH, United Kingdom}
\author{R.~H.~Maruyama}
\affiliation{Department of Physics and Wright Laboratory, Yale University, New Haven, Connecticut 06520, USA}
\author{F.~Mouton}
\affiliation{Department of Physics and Astronomy, University of Sheffield, Sheffield S3 7RH, United Kingdom}
\author{S.~L.~Olsen}
\affiliation{Center for Underground Physics, Institute for Basic Science (IBS), Daejeon 34126, Republic of Korea}
\author{B.~J.~Park}
\affiliation{IBS School, University of Science and Technology (UST), Daejeon 34113, Republic of Korea}
\author{H.~K.~Park}
\affiliation{Department of Accelerator Science, Korea University, Sejong 30019, Republic of Korea}
\author{H.~S.~Park}
\affiliation{Korea Research Institute of Standards and Science, Daejeon 34113, Republic of Korea}
\author{K.~S.~Park}
\affiliation{Center for Underground Physics, Institute for Basic Science (IBS), Daejeon 34126, Republic of Korea}
\author{R.~L.~C.~Pitta}
\affiliation{Physics Institute, University of S\~{a}o Paulo, 05508-090, S\~{a}o Paulo, Brazil}
\author{H.~Prihtiadi}
\affiliation{Department of Physics, Bandung Institute of Technology, Bandung 40132, Indonesia}
\author{S.~J.~Ra}
\affiliation{Center for Underground Physics, Institute for Basic Science (IBS), Daejeon 34126, Republic of Korea}
\author{C.~Rott}
\affiliation{Department of Physics, Sungkyunkwan University, Suwon 16419, Republic of Korea}
\author{K.~A.~Shin}
\affiliation{Center for Underground Physics, Institute for Basic Science (IBS), Daejeon 34126, Republic of Korea}
\author{A.~Scarff}
\affiliation{Department of Physics and Astronomy, University of Sheffield, Sheffield S3 7RH, United Kingdom}
\author{N.~J.~C.~Spooner}
\affiliation{Department of Physics and Astronomy, University of Sheffield, Sheffield S3 7RH, United Kingdom}
\author{W.~G.~Thompson}
\affiliation{Department of Physics and Wright Laboratory, Yale University, New Haven, Connecticut 06520, USA}
\author{L.~Yang}
\affiliation{Department of Physics, University of Illinois at Urbana-Champaign, Urbana, Illinois 61801, USA}
\author{G.~H.~Yu}
\affiliation{Department of Physics, Sungkyunkwan University, Suwon 16419, Republic of Korea}
\collaboration{COSINE-100 Collaboration}